# Cataclysmic Variables from USNO-B1.0 Catalog: Stars with Outbursts on Infrared Palomar Plates

D. V. Denisenko*

*Space Research Institute, Moscow, Russia*

**Abstract**–The search was performed for cataclysmic variable stars with outlying infrared magnitudes in USNO-B1.0 catalog. The selection was limited to objects in the Northern hemisphere with ($B1$-$R1$) < 1 and ($R2$-$I$) > 3.5. The search method is described, and the details on individual stars are given. In total 27 variable objects were found with 20 being the previously known ones and 7 – new discoveries. 4 of newly found variables are dwarf novae, while the remaining 3 – pulsating red variables of Mira or semi-regular types, including a heavily reddened one in Pelican nebula. The correct pulsation period is determined for two stars discovered by other researchers, and the dwarf nova nature is confirmed for another object previously suspected as such. The perspectives of the proposed search technique in the framework of Virtual Observatory are also discussed.

Key words: *stars, cataclysmic variables*

INTRODUCTION

Cataclysmic variables are a class of stellar systems deserving a special astrophysical interest. These objects provide important information for understanding the evolution of binary systems and are considered to be the progenitors of Supernovae of Ia type. The search for new variables of this type provides the better statistics of the stellar population in our galaxy and in some cases brings up the new objects with unusual properties. This is why new cataclysmic variables are typically attracting the attention of researchers, especially the systems bright enough for studying with the moderate class telescopes. Objects brighter than 17-18$^m$ are suitable for time resolved photometry and spectroscopy, for measuring the principal parameters of the binary system such as orbital period, inclination, component masses and even sizes in case of the eclipsing variable. Every new CV improves our knowledge base and helps to obtain the better distributions of various parameters, to check the theoretical models and heuristic dependencies.

Still the cataclysmic variables are often discovered serendipitously, usually as the optical transients during the dwarf nova outbursts. Many CVs were discovered this way in the last few years by Catalina Real-Time Transient Survey (CRTS), see Drake et al., 2009. In the same time the systematic search for CVs has been performed in the data from Sloan Digital Sky Survey (SDSS) based on the spectra and multi-color photometry of millions of objects. The results from SDSS are less biased compared to CRTS, but the Sloan Survey is only covering a small fraction of the sky. Catalina Sky Survey is also avoiding the dense regions of Milky Way where the large number of cataclysmic variables resides. Perhaps the only uniform database covering the whole celestial sphere is Digitized Sky Survey based on the photographic plates taken at the Palomar observatory in the Northern hemisphere and Anglo-Australian observatory in the Southern skies. The images taken in three photometric bands (Blue, Red and Infrared) at different epochs were used as a source for the USNO-B1.0 catalog published in 2003 (Monet et al.). This article is an attempt to search for the cataclysmic variables in USNO-B1.0 and to evaluate the perspectives of this approach within the concept of Virtual Observatory.

---

* E-mail: denis@hea.iki.rssi.ru

# SEARCH METHOD

The search method used in this work is similar to the one used for identification of cataclysmic variables in 1RXS catalog (Denisenko and Sokolovsky, 2011). Suspected objects from USNO-B1.0 catalog (Monet et al., 2003) with white *B-R* index and the large difference between *I* and *R* magnitudes are selected for checking on the digitized Palomar Sky Survey (DSS) plates. The main difference is that now it is not limited to 30″ circles around the ROSAT X-ray sources, but covers the entire sky. Due to the huge amount of stars in USNO-B1.0 catalog (1.046 billion objects) the search criteria were tightened compared to the 1RXS article. The variable stars were searched among the objects which obeyed the following conditions:

- *B*1, *R*1, *B*2, *R*2 and *I* magnitudes are all present in USNO-B1.0 (formally this is coded as B1*R1*B2*R2*I <> 0);
- *B*1-*R*1 < 1 (to exclude red variables from the search as much as possible);
- *R*2-*I* > 3.5 (to find the stars with outbursts on IR plates);
- Declination $\delta > 0$ (to use the uniform data obtained with the same Schmidt camera).

The previous experience has shown that the "Star/Galaxy Separators" in USNO-B1.0 can not be used reliably to distinguish stars from compact galaxies. For some stars which were known to be genuine CVs the catalog was giving S/G values as low as 3 or 4 (formally a galaxy) for *R*1 and *B*2 bands, but 8 or 10 (typical for stars) for *R*2, *B*1 and *I* bands. In other cases, however, *R*1 and *B*2 separators were higher than 8, with *R*2 and *B*1 S/G being small. On the contrary, some faint round galaxies had typically stellar values of S/G separators (8, 9 and even 10) which made it impossible to distinguish them from stars. Thus the S/G separators were ignored at all, and the final decision was made after the visual inspection of the scanned Palomar plates.

The simple code was written to extract the candidate variable blue stars from the catalog, and the search has started from the North Pole towards the celestial equator. This is why the stars found during the search are listed in this paper in the order of decreasing declinations. In total, over 7000 candidate objects were automatically selected. A large amount of objects were apparently "false alarms" which was obvious from their concentration towards the plate edges where the large number of plate defects reside. Such clusters were excluded from the following study.

The remaining 5060 candidates were checked visually on the scanned Palomar plates available online at DSS Plate Finder website http://archive.stsci.edu/cgi-bin/dss_plate_finder. This site allows downloading images from all available plates for selected area, while the form at http://archive.stsci.edu/cgi-bin/dss_form only shows one default image for each band. For some areas of sky the default plate is not the same one from which the magnitudes for USNO-B1.0 catalog were measured. The search area was selected to be 5′×5′, the file format used was GIF. In special cases when the image quality in browser was not good enough the original FITS files were downloaded and examined more carefully in MaxIm DL software, aligning images when necessary and matching screen stretch of images obtained at different epochs.

The results of checking 5060 candidate objects with (*B*1-*R*1) < 1 and (*R*2-*I*) > 3.5 are as follows. 21 "objects" are due to the satellite trails crossing infrared images of real, but faint stars. 309 are airplane trails/wing light flares superimposed on the stellar image. 1 is a meteor crossing the star. 105 are various plate defects. At least 259 objects turned to be galaxies (diffuse objects). 330 "objects" are artifacts caused by diffraction spikes/circles from the nearby bright stars. In 4 cases there are nearby flares of stellar appearance very close to the faint star position, but not exactly coincident with it. 1 "false alarm" is caused by the track of asteroid (311) Claudia which was occulting a faint star during the exposure. 513 objects are close pairs of stars and 73 – groups of 3 and more stars which were mixed up with each other due to either proper motion between 1$^{st}$ and 2$^{nd}$ Palomar epoch or due to imperfect astrometric solutions, especially near the plate edges. 3340 catalog entries with ($R2$-$I$) > 3.5 are not real without any apparent reason. Most likely they are explained by the bad zero point calibration in one or both photometric band(s). Finally, 27 candidates are real variable objects (20 previously known and 7 new ones).

The objects found to be variable were studied in more detail. All available DSS plates of 15′×15′ areas centered at the new object were downloaded in FITS format. If there are archival NEAT images, they were also downloaded and analyzed. Two objects were also checked on the archival photographic plates of Moscow collection at SAI MSU. The results of the search and detailed analysis of individual objects of special interest are presented in the next two chapters.

KNOWN VARIABLE OBJECTS

As mentioned above, the search was aimed at finding the cataclysmic variables which are blue compared to the majority of stars in our galaxy. Since the blue and red Palomar plates of the 1$^{st}$ epoch were taken on the same night, immediately one after another, small ($B1$-$R1$) color index of an object in USNO-B1.0 should, in principle, point at the intrinsically blue objects. Combined with the large value of $R2$-$I$ (3.5 or higher) this should limit the search results to the dwarf novae and active quasars with outbursts on infrared plates. Indeed, 10 objects of 20 among the known variables were blue, 7 of them being dwarf novae and 3 – QSO (see Table 1). However, 10 other objects found during the search turned out to be pulsating red variables of Mira type. Even more, three of them had negative values of ($B1$-$R1$) color index: V418 Cas (-0.85), CZ Her (-0.52) and CY Sge (-0.62). The analysis of plates has shown that the variable stars were at minimum light during the imaging. Since those stars are located in the rich parts of the Milky Way, the light of the nearby field stars was measured instead of Mira during the digitization of the corresponding plates. In some cases it is also possible that the blue ($B1$) magnitude was measured of the plate from one standard area, and the red ($R1$) – of the plate from the neighboring area, should it have a better image quality near the plate edge.

**Table 1.** Previously known variable objects

| RA | Dec | $B1$ | $R1$ | $B2$ | $R2$ | $I$ | Name | Type |
|---:|---:|---:|---:|---:|---:|---:|---|---|
| 154.554292 | 71.928764 | 18.58 | 18.44 | 19.13 | 19.39 | 14.42 | CI UMa | DN |
| 37.469069 | 69.281847 | 14.75 | 14.43 | 14.02 | 13.96 | 9.29 | NSV 832 | |
| 18.723317 | 63.612439 | 15.07 | 14.14 | 13.32 | 14.39 | 9.90 | V724 Cas | M |
| 18.249397 | 62.179792 | 17.53 | 18.38 | 17.14 | 15.98 | 10.20 | V418 Cas | M |
| 129.178058 | 53.477269 | 16.40 | 16.27 | 16.81 | 16.39 | 11.74 | SW UMa | DN |
| 24.244106 | 47.858172 | 19.11 | 18.63 | 17.18 | 19.25 | 15.21 | QSO B0133+476 | QSO |
| 50.063650 | 44.183331 | 19.35 | 18.59 | 18.70 | 18.71 | 15.01 | SDSS J032015.29+441059.3 | DN |
| 349.900250 | 36.783406 | 17.73 | 17.73 | 18.66 | 17.73 | 13.63 | 1RXS J231935.0+364705 | DN |
| 117.248117 | 31.420206 | 18.92 | 17.94 | 18.18 | 18.72 | 14.69 | SDSS J074859.55+312512.6 | DN |
| 179.882678 | 29.245547 | 17.33 | 16.39 | 17.45 | 17.39 | 13.39 | QSO 4C 29.45 | QSO |
| 239.035064 | 29.238347 | 10.91 | 10.86 | 11.81 | 12.94 | 9.02 | Z CrB | M |
| 271.36098 | 22.501086 | 13.79 | 14.31 | 12.52 | 13.84 | 9.43 | CZ Her | M |
| 304.611253 | 19.132556 | 14.86 | 15.48 | 14.80 | 14.95 | 10.63 | CY Sge | M |
| 297.267108 | 15.423714 | 14.18 | 13.35 | 12.58 | 14.15 | 10.00 | OS Aql | M |
| 280.958042 | 15.302433 | 17.40 | 16.97 | 15.53 | 16.46 | 11.29 | NSVS J1843498+151805 | M |
| 269.491206 | 15.016214 | 14.27 | 13.67 | 15.39 | 14.14 | 10.12 | NSVS J1757576+150053 | M |
| 137.864103 | 8.694686 | 19.85 | 19.42 | 20.20 | 19.27 | 15.65 | SDSS J091127.36+084140.7 | DN? |
| 340.951992 | 8.157508 | 18.10 | 18.74 | 19.03 | 16.62 | 13.08 | ASAS J224349+0809.5 | DN |
| 298.639903 | 5.682869 | 14.17 | 14.16 | 15.09 | 12.21 | 8.12 | V1062 Aql | M |
| 34.454011 | 1.747183 | 18.52 | 18.72 | 19.89 | 18.51 | 13.29 | QSO J0217+0144 | QSO |

Several objects deserve a special attention and are discussed below.

**SDSS J091127.36+084140.7** was identified as a CV from SDSS spectrum which shows bright Balmer emission lines on the blue continuum, but no He lines (Szkody et al., 2005). Follow-up time-resolved spectroscopy with VLT (Southworth et al., 2006) has shown the orbital period to be 0.2054d. The star is listed in VSX (Watson et al., 2006) and RKcat (Ritter and Kolb) as possible dwarf nova (DN?) since no outburst has been reported in literature. Finding the star in outburst on Palomar IR plate taken on 1996 Jan. 13 confirms the dwarf nova classification for this variable. It is faint on the remaining eight DSS plates. In addition, CRTS archival light curve shows 4 outbursts on 2007 Feb. 15 (to 15.3$^m$), 2007 Nov. 18, 2008 Feb. 14 and 2011 Jan. 02 (all to 16.0$^m$). Rather long orbital period is favoring the UGSS subtype for J0911+0841.

**OS Aql**, despite being a Mira, was originally thought to be a dwarf nova and was even included in 1957 Atlas of Variable Stars of U Gem type (Brun and Petit). 2MASS color index ($J-K$=1.89) confirms its Mira nature without any doubt.

**NSVS J1843498+151805** is listed in VSX as Mira with a period of 212d based on NSVS data (Wozniak et al., 2004) with other name V827 Her which is a classical nova (Nova Her 1987). Analysis of Palomar plates definitely shows they are two different variable stars separated by 2′. Variability range for NSVS J1843498+151805 from ROTSE-I data is quoted to be 11.39-13.33. I have checked the ASAS-3 data (Pojmanski, 1997) at http://www.astrouw.edu.pl/asas/ and have found the object ASAS184349+1518.2 which is the same as NSVS J1843498+151805. The variable is present in 2MASS catalog as 2MASS 18434993+1518087 with infrared magnitudes $J$=7.728±0.023, $H$=6.746±0.044, $K$=6.234±0.017 typical for Mirae. The maximum $V$ magnitude of ASAS184349+1518.2 is 13.61, the minimum is well below the ASAS limiting magnitude (15.2). Comparison of two blue Palomar plates shows that the amplitude of variability is actually more than 5 magnitudes. Star is bright on 2$^{nd}$ epoch plate taken on 1990 June 21 ($B$=15.53), but nothing is visible at the position of variable on 1$^{st}$ epoch plate of 1951 July 12 (limiting magnitude $B$~20.5). See Figure 1 for comparison of two blue Palomar plates.

**NSVS J1757576+150053** is listed in VSX as Mira with a period of 348d based on NSVS data (Wozniak et al., 2004). Infrared magnitudes of 2MASS 17575790+1500582 ($J$=7.056±0.034, $H$=6.142±0.016, $K$=5.615±0.017) confirm the Mira classification. I have checked ASAS-3 data and have found the object ASAS175758+1501.0 which is the same as NSVS J1757576+150053. The light curve of ASAS175758+1501.0 covering 7 years is shown in Figure 2. ASAS data were analyzed using WinEffect period search software by V. P. Goranskij. The period was found to be 318±2d, not consistent with the findings of Wozniak et al. Obviously NSVS data were covering too short interval of time (about a year) for the correct period determination of this variable.

**V1062 Aql** is listed in GCVS and VSX as a Mira variable, but without a pulsation period. I have checked ASAS-3 data for ASAS195434+0540.9 covering 7 years of observations. Four maxima were observed by ASAS with two more missed due to the seasonal gap. The best value of period I have determined from ASAS data is 505±2d. It is also important to note that the variable is located 2″ north-west of another field star which is several magnitudes brighter than V1062 Aql itself at minimum light, but fainter during the maximum. The comparison of blue Palomar plates taken on 1953 Sep. 05 and on 1995 July 31 is shown in Figure 3.

Variability of **1RXS J231935.0+364705** was found by the author on 2007 Dec. 08 (Denisenko, 2009) using NSVS data. Two outbursts to $R$~14.5 have been detected on 2009 Nov. 05 and 2010 Dec. 26, but the orbital period has not been determined yet. **SDSS J032015.29+441059.3** and **SDSS J074859.55+312512.6** were identified as dwarf novae by Wils et al. in 2009. **ASAS**

**J224349+0809.5** was identified by P. Wils on 2009 Sep. 22 and observed in outburst on 2009 Oct. 06 (Shears et al., 2010). All these interesting variables could have been discovered at least 5 years earlier from USNO-B1.0 catalog data alone.

NEW VARIABLE STARS

Seven variable objects found during this work were not present in GCVS (Samus et al.), Simbad and VSX as of 2011 Aug. 25, and no mention of their variability was found on the web using Google search for coordinates and names. Following the numbering scheme introduced back in 2007, these newly discovered variables were designated DDE 22 - DDE 28, using AAVSO observer's code for the author. The details on these objects, including their USNO-B1.0 magnitudes, are given in Table 2. The list of variables discovered by DDE with their coordinates, finder charts and classification is available online at http://hea.iki.rssi.ru/~denis/VarDDE.html.

**Table 2.** Newly discovered variables in USNO-B1.0 catalog

| Name | USNO-B1.0 | RA | Dec | $B1$ | $R1$ | $B2$ | $R2$ | $I$ | Type |
|---|---|---|---|---|---|---|---|---|---|
| DDE 22 | 1597-0071405 | 67.67893 | 69.74123 | 17.78 | 17.59 | 18.66 | 19.79 | 15.88 | DN |
| DDE 23 | 1481-0306480 | 274.888 | 58.10732 | 19.02 | 19.06 | 20.27 | 19.88 | 16.37 | DN |
| DDE 24 | 1338-0387240 | 312.1372 | 43.85539 | 18.70 | 17.89 | 18.06 | 18.70 | 14.87 | M: |
| DDE 25 | 1272-0325825 | 263.696 | 37.24594 | 20.17 | 19.20 | 20.27 | 20.41 | 16.58 | DN |
| DDE 26 | 1209-0561869 | 330.8676 | 30.94355 | 19.38 | 18.73 | 20.06 | 19.25 | 15.39 | DN |
| DDE 27 | 1058-0501589 | 296.8936 | 15.81273 | 16.55 | 15.58 | 16.41 | 14.17 | 9.83 | M |
| DDE 28 | 1024-0485702 | 286.6826 | 12.45965 | 17.50 | 16.56 | 17.95 | 16.91 | 13.31 | M |

Table 3 below lists coordinates of new variable stars in hhmmss ddmmss format, their names and magnitudes in USNO-A2.0 catalog, variability range and constellation to which they belong.

**Table 3.** Coordinates and USNO-A2.0 designations of new variables

| Name | RA (2000.0) | Dec (2000.0) | USNO-A2.0 | $R1$ | $B1$ | Range, $R$ | Const |
|---|---|---|---|---|---|---|---|
| DDE 22 | 04 30 42.94 | +69 44 28.4 | 1575-02049484 | 17.7 | 16.7 | 15.9–19.8 | Cam |
| DDE 23 | 18 19 33.12 | +58 06 26.3 | 1425-09097975 | 18.6 | 18.8 | 16.4–20.: | Dra |
| DDE 24 | 20 48 32.92 | +43 51 19.4 | 1275-14170832 | 14.8 | 17.8 | 14.8–18.8 | Cyg |
| DDE 25 | 17 34 47.03 | +37 14 45.4 | 1200-08493043 | 16.1 | 15.9 | 16.1–20.: | Her |
| DDE 26 | 22 03 28.21 | +30 56 36.8 | 1200-18847099 | 18.7 | 19.0 | 15.4–19.3 | Peg |
| DDE 27 | 19 47 34.46 | +15 48 46.8 | 1050-14926876 | 15.7 | 16.6 | 15.7–<20 | Aql |
| DDE 28 | 19 06 43.83 | +12 27 34.7 | 0975-14117111 | 14.7 | 18.4 | 14.7–<20 | Aql |

**Note.** Real range of variability may be well beyond the values shown in this table.

New variables are discussed in more detail below.

**DDE 22.** The star was in outburst on 1997 Nov. 07 infrared plate and at quiescence on all five remaining plates. Comparison of red and infrared Palomar plates is presented in Figure 4. The star is listed in GALEX catalog as GALEX J043042.9+694428 with the far and near ultraviolet magnitudes FUV=20.68±0.19, NUV=20.71±0.15 and in 2MASS as 2MASS 04304278+6944281 with infrared magnitudes $J$=16.562±0.129, $H$=16.343±0.245, $K$=N/A. UV and IR colors together with the proper motions from USNO-B1.0 catalog (-4 and -6 mas/yr in Right Ascension and in Declination, respectively) confirm the cataclysmic variable nature of this object.

The position of this star is inside the field of view of photographic plates centered at BU Cam which were taken at Crimean laboratory of SAI MSU in 1989-1991. 40 plates stored in Moscow collection were visually checked. Star was found in bright outbursts ($B$=15.6) on plates taken on 1990 Sep. 18, Sep. 21 and Oct. 28.

**DDE 23.** The star was in outburst on 1996 July 16 infrared plate and at quiescence on all eleven remaining plates. Comparison of red plate taken on 1994 July 10 and infrared plate of 1996 July 16 is presented in Figure 5. The star is listed in GALEX catalog as GALEX J181933.1+580625 with the far and near ultraviolet magnitudes FUV=19.19±0.10, NUV=19.13±0.07 and in 2MASS as 2MASS 18193311+5806258 with infrared magnitudes $J$=16.235±0.111, $H$=15.601±0.133, $K$=15.321±0.147. UV and IR colors together with the proper motions from USNO-B1.0 (-8, -12 mas/yr) confirm the cataclysmic variable nature of this object.

**DDE 24.** Variable is located within the dark region at the south-western part of Pelican nebula (IC 5070) with a huge interstellar absorption. Comparison of blue Palomar plates taken on 1989 July 10 and 1993 June 27 is presented in Figure 6. Photometry of DDE 24 from Palomar plates using USNO-A2.0 1275-14171682 (R.A.=20 48 35.46, Dec.=+43 53 13.8, $R$=15.7, $B$=17.4) as a comparison star is given in Table 4.

**Table 4.** Photometry of DDE 24 from red and blue Palomar plates

| Date | JD | $R$ mag | Date | JD | $B$ mag |
|---|---|---|---|---|---|
| 1953 June 14 | 2434542.928 | 16.78 | 1953 June 14 | 2434542.907 | 19.91 |
| 1954 July 05 | 2434928.894 | 15.47 | 1954 July 05 | 2434928.899 | 17.57 |
| 1990 Sep. 11 | 2448145.728 | 17.28 | 1989 July 10 | 2447717.847 | 19.37 |
| 1992 Aug. 27 | 2448861.727 | 17.28 | 1993 June 27 | 2449165.910 | 17.69 |

The variable is present in 2MASS catalog as 2MASS 20483292+4351190 with an enormously high ($J$-$K$) color index or 3.42±0.03: $J$=13.411±0.026, $H$=11.624±0.027, $K$=9.993±0.019. Comparison of $J$ and $K$ 2MASS images of DDE 24 obtained on 2000 May 24 is presented in Figure 7. Note the approximately 2′×2′ region around the variable where there are no stars at all on the $J$-band image ($J$>17.0), while the $K$-band image has 4 additional stars up to $K$=13.2. Given quite a moderate $B$-$R$ color of the variable on 1954 July 05, it is possible that DDE 24 is actually a star moving behind the knots of the dark nebula, sometimes showing itself through the less dense parts of a cloud, and its variability may be caused purely by the external reasons.

The position of this star falls inside one of the Standard Areas (SA 40) imaged with the Crimean astrograph of SAI MSU. 80 plates from Moscow collection were visually inspected. The star is below the limit on all of them. Typical limiting magnitude of the Moscow plates is $B$=16.5-17.

**DDE 25.** Two outbursts were detected on 3 Palomar plates: IR plate of 1994 Apr. 20, as well as blue and red POSS-I plates taken on 1951 July 08. Comparison of two 1st epoch blue plates taken on 1951 July 08 and 1954 July 05 is shown in Figure 8. It is remarkable that USNO-A2.0 magnitudes were measured off 1951 plates, this is why the variable is included into this catalog as USNO-A2.0 1200-08493043 with the magnitudes $R$=16.1, $B$=15.9. USNO-B1.0 magnitudes listed in Table 2 are obviously from 1954 plates.

The star has three entries in GALEX catalog with the colors compatible with CV classification:

GALEX J173447.0+371446, FUV=21.73±0.44, NUV=20.89±0.23
GALEX J173447.0+371444, FUV=20.43±0.19, NUV=20.72±0.19
GALEX J173447.1+371443, FUV=20.57±0.26, NUV=20.57±0.19

**DDE 26.** Star is in outburst on two infrared Palomar plates taken on 1992 July 22 and 23 and at quiescence on seven other plates. It is not present in 2MASS catalog. Position of DDE 26 falls just outside the GALEX field of view, thus UV magnitudes are not available, too. And there is no proper motion measured in USNO-B1.0 catalog. Luckily, the area of this object was observed by Palomar/NEAT project (Teegarden et al., 2003) on five nights in 2001-2002. Fifteen images were taken, and on three images obtained on 2002 July 28 the star was in outburst. Comparison of combined NEAT images of DDE 26 obtained in outburst and at quiescence is presented in Figure 9. The light curve from NEAT data is shown in Figure 10.

**DDE 27.** Star was at minimum light on 1950 July 17 when the POSS-I plates were taken. This is why the USNO-B1.0 coordinates and magnitudes were actually measured for the nearby star about 1.5″ away (R.A.=19 47 34.468, Dec.=+15 48 45.83). The variable itself is a red star 2MASS 19473447+1548444 (R.A.=19 47 34.475, Dec.=15 48 44.49) with the following magnitudes: $J$=8.724±0.023, $H$=7.779±0.034, $K$=7.213±0.027. It is too faint for ASAS.

**DDE 28.** This star was bright on POSS-I plates of 1952 May 24, but faded below the blue plate limit ($B$<20.5) by 1952 Aug. 12. It is identical to the infrared object 2MASS 19064388+1227347 (R.A.=19 06 43.884, Dec.=+12 27 34.79, $J$=9.586±0.020, $H$=8.404±0.020, $K$=7.880±0.016).

DISCUSSION

As shown in this work, many objects with large amplitude of variability remain undiscovered for years and sometimes even decades, despite being obvious on archival plates and in astrometric catalogs. A number of those objects have a real astrophysical importance. Some variable stars discovered as optical transients have later proven to be not what they were originally thought. For example, 1RXS J184542.4+483134 tentatively identified as a dwarf nova turned out to be the magnetic CV (polar) with unprecedented large amplitude between low and high states of ~5$^m$ and a short period 0.05491d (Denisenko and Smirnov, 2011; Pavlenko et al., 2011). Since these objects are being only discovered recently, the precious time may be lost that could have provided the critical long-term observational data necessary for understanding their nature.

The search presented in this work was limited to the objects with ($B$1-$R$1) < 1 and ($R$2-$I$) > 3.5 only. Although 27 real variable objects were identified which is not enough for statistically confident conclusions, the results can help in estimating the number of variables yet to be found in Palomar plates and USNO-B1.0 catalog. Seven new variables were found in this work, with five more cataclysmic variables discovered by other methods in 2005-2010, after the USNO-B1 catalog was published. The search for stars with outbursts on one of blue or red plates carried out by the author has resulted in discovery of two cataclysmic variables in Lyra, DDE 20 and DDE 21 (Denisenko, 2011). DDE 20 was found due to the value of $R$1-$R$2=3.27, while DDE 21 has an outburst on the infrared plate to $I$=11.75. This star would not have been found using the formal requirements in this work since it has $B$1-$R$1 color index larger than 1.0 ($B$1=20.48, $R$1=19.14). Many cataclysmic variables, especially the recurrent novae with the large contribution from the secondary companion, have intrinsically red magnitudes at quiescence. Extending the search to the objects with |$R$1-$R$2|>2.5, |$B$1-$B$2|>2.5 and |$R$2-$I$|>2.5 one can expect to identify a few dozen additional dwarf novae with outbursts on the Palomar plates. Obviously a number of candidates will dramatically increase to the amount not manageable by one researcher within a reasonable time. Author's personal experience shows that 5000 candidates can be checked in 250 man-hours of work, that is, in a month entirely dedicated to blinking of the scanned Palomar plates eight hours per day. Loosening the "outburst threshold" to 3 or even to 2.5 will result in approximately 20,000 and 80,000 candidates requiring thousands man-hours. Perhaps this task can be achieved by using the resources of amateur community ("crowdsourcing"). The search can be extended to the Southern hemisphere and to the objects with one or two magnitudes missing in USNO-B1.0.

Another direction for the future search is using two or three catalogs together, for example, correlating blue and red USNO-B1.0 magnitudes with those from USNO-A2.0 and even with *V* magnitudes from GSC 2.3.2. This task can be performed using the Virtual Observatory tools. Work in this direction has started at the Russian Institute of Problems of Informatics together with astronomers from the Space Research Institute and Sternberg Astronomical Institute of Moscow State University.

## CONCLUSION

Using the magnitudes from USNO-B1.0 catalog only, I have discovered seven new variable stars in the Northern hemisphere, including four new cataclysmic variables and one heavily reddened variable in Pelican nebula. This work once again shows that it is possible to find new objects of astrophysical interest without making a single observation, based on the archival images and catalogs. The follow-up monitoring of the newly discovered variables, as well as the continued search for stars with the high amplitude of variability is encouraged. Efforts and ideas on the joint work are also welcome.

## ACKNOWLEDGMENTS

I would like to thank E. Gorbovskoy and A. Belinski (SAI MSU) for providing the copy of USNO-B1.0 catalog, N. N. Samus (SAI MSU) for his permission to work with the Moscow photographic plate archive, S. V. Antipin, E. V. Kazarovets and E. N. Pastukhova (SAI MSU) for helpful discussions. I thank the Program of support of the leading scientific schools (grant NSh-5069.2010.2). This work has made an intensive use of digitized plates from the Palomar Observatory Sky Survey obtained by Caltech and supported by NSF, NGS, Sloan Foundation, Samuel Oschin Foundation and Eastman Kodak Co. While preparing this article I have made use of the SIMBAD database operated at CDS, Strasbourg, France, and the Variable Star Index (VSX) maintained by the American Association of Variable Star Observers.

FIGURES

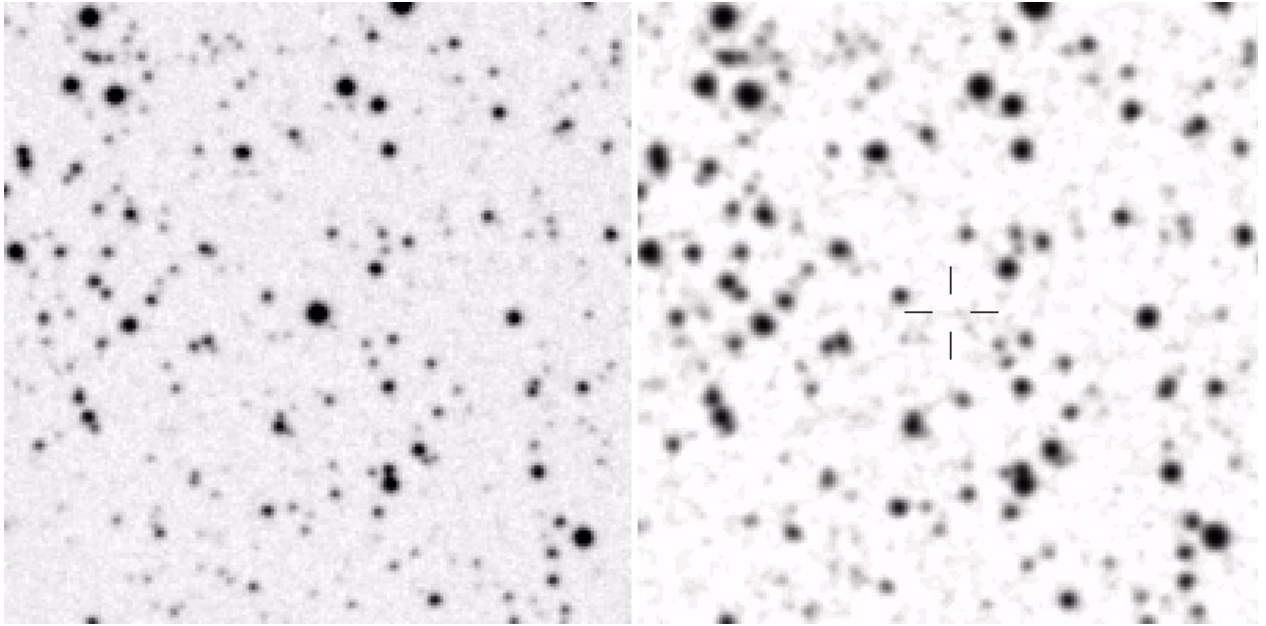

Figure 1. 200″×200″ finder chart of NSVS J1843498+151805. Left: Blue Palomar plate taken on 1990 June 21. Right: Blue Palomar plate of 1951 July 12. North is up, East is to the left.

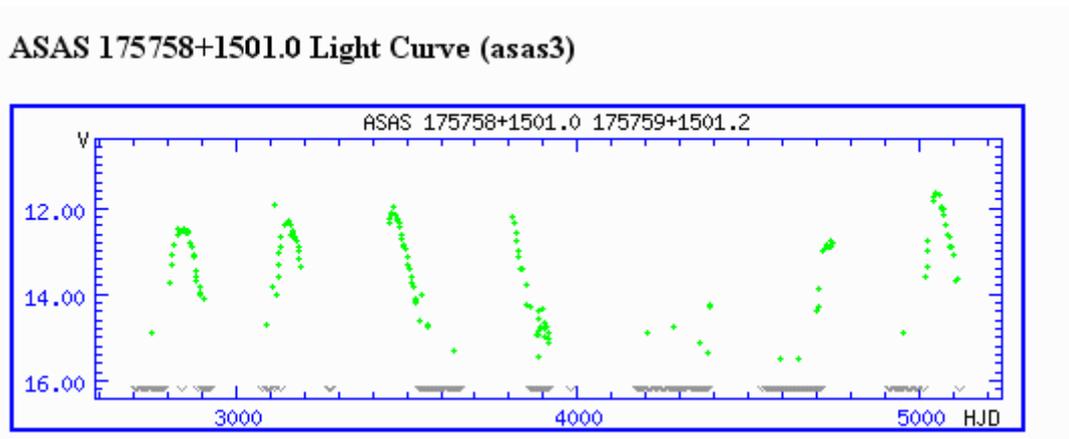

Figure 2. Light curve of ASAS175758+1501.0 covering 7 years.

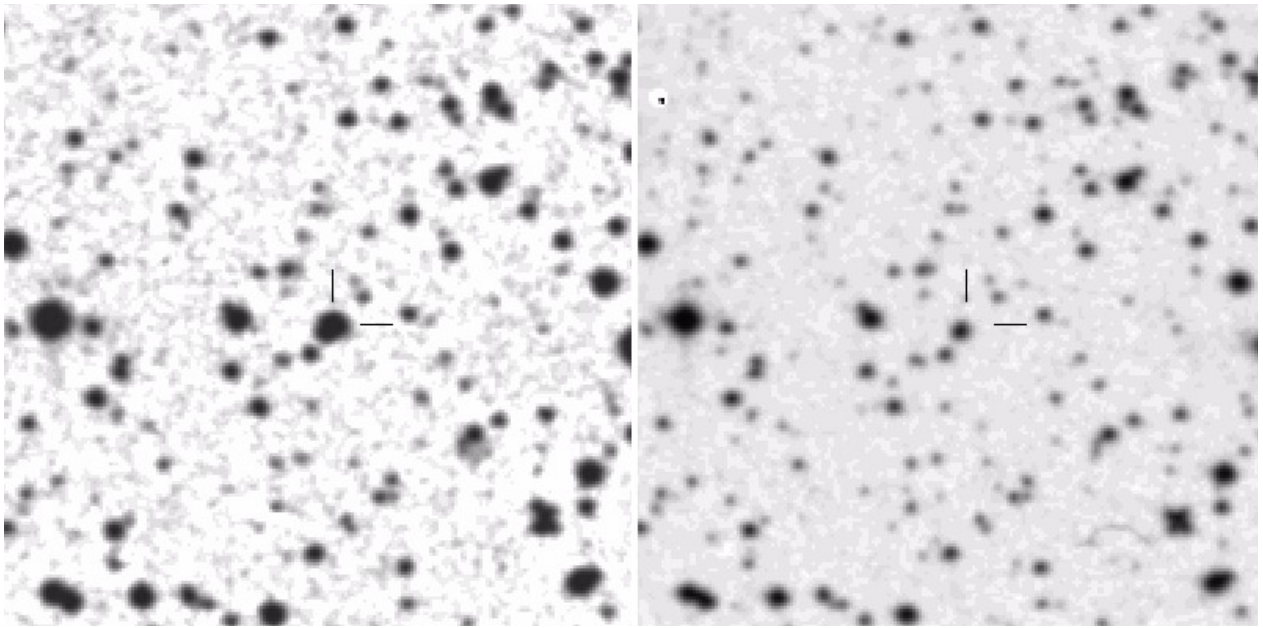

Figure 3. Palomar images of V1062 Aql. Left: Blue plate taken on 1953 Sep. 05. Right: Blue plate of 1995 July 31. Field of view is 200″×200″.

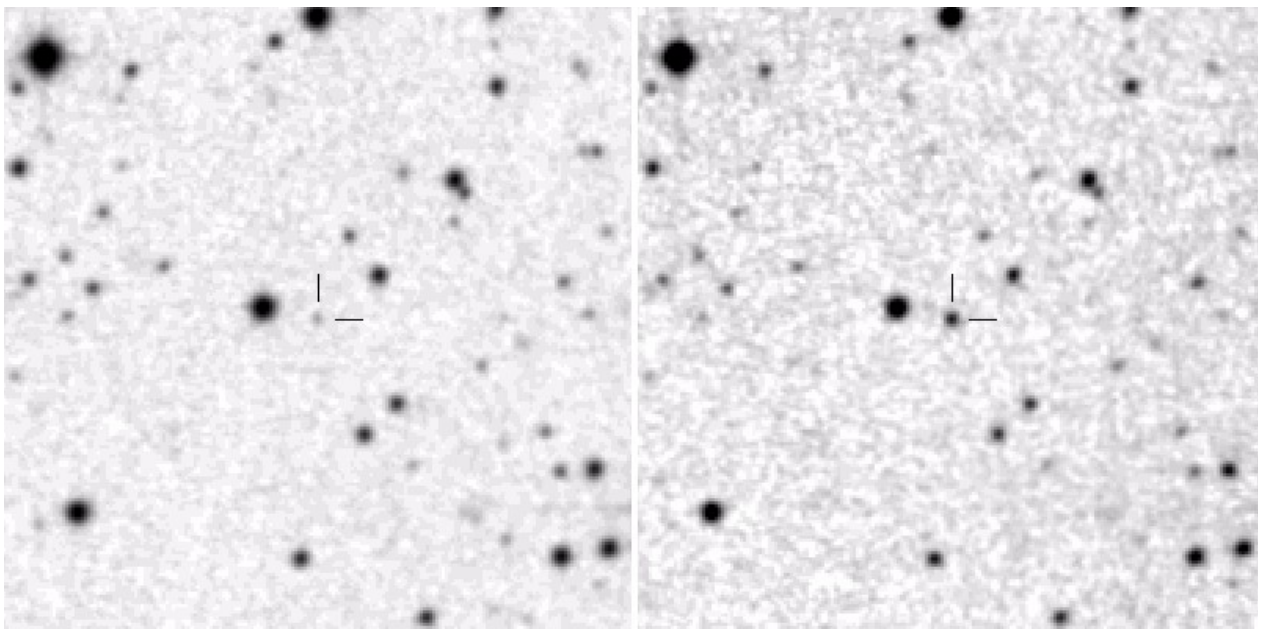

Figure 4. Palomar images of DDE 22. Left: red plate taken on 1993 Sep. 21. Right: infrared plate of 1997 Nov. 07.

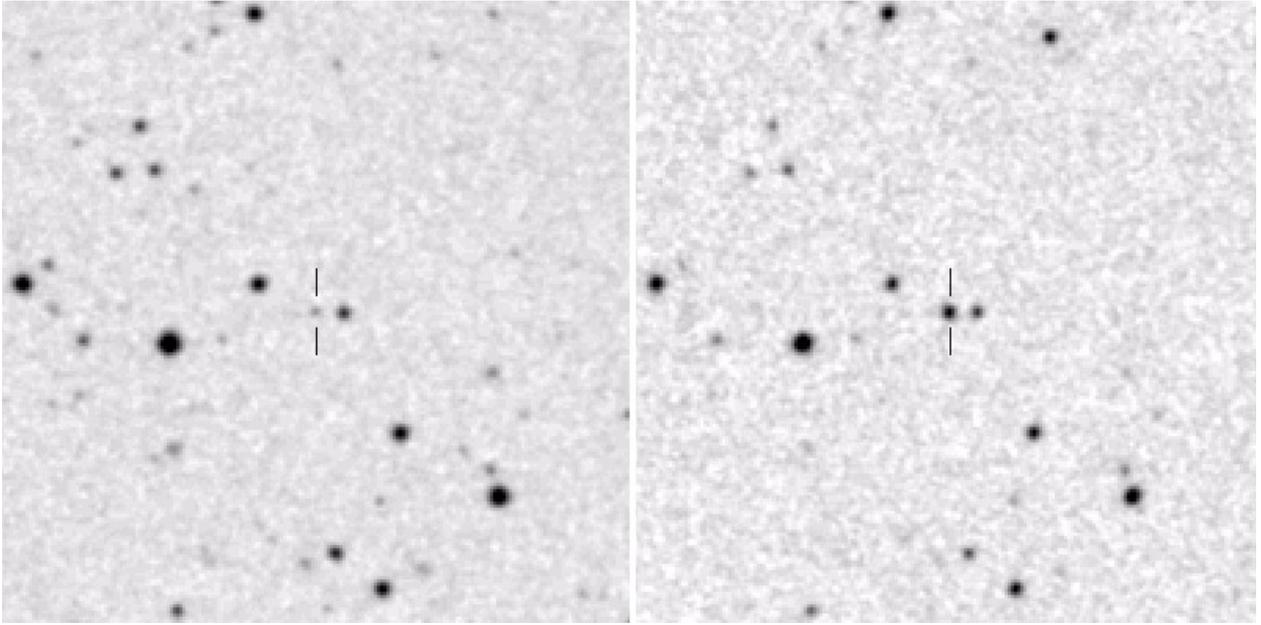

Figure 5. Palomar images of DDE 23. Left: red plate taken on 1994 July 10. Right: infrared plate of 1996 July 16.

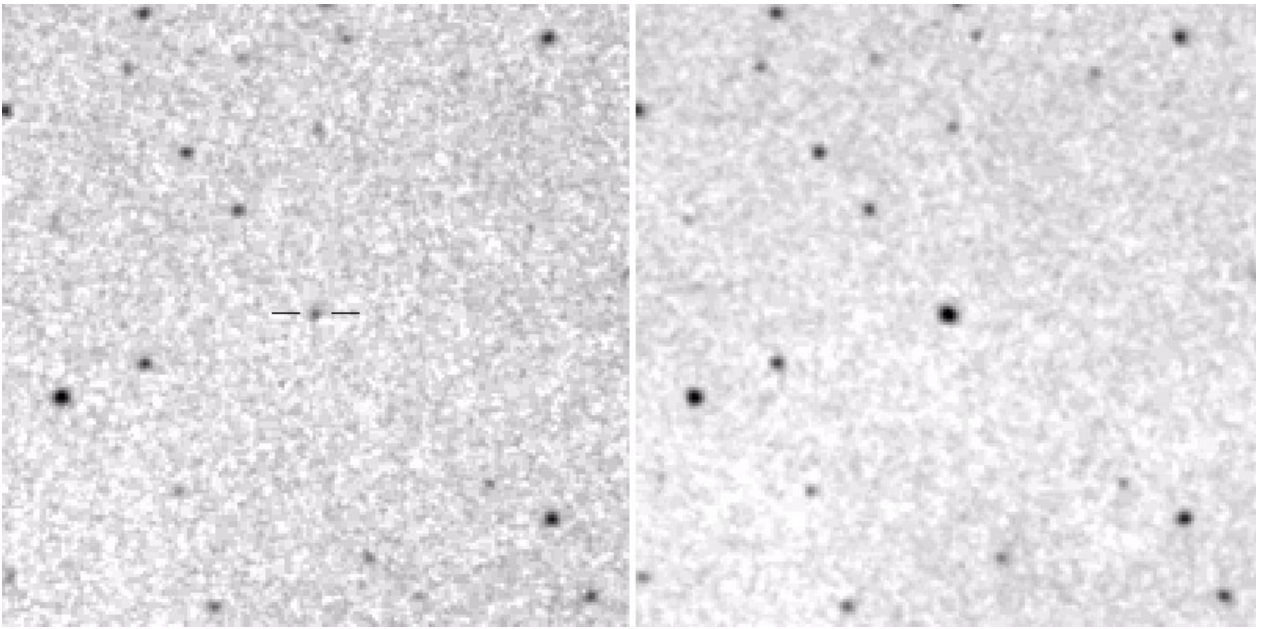

Figure 6. Blue Palomar images of DDE 24. Left: 1989 July 10 plate. Right: 1993 June 27 plate.

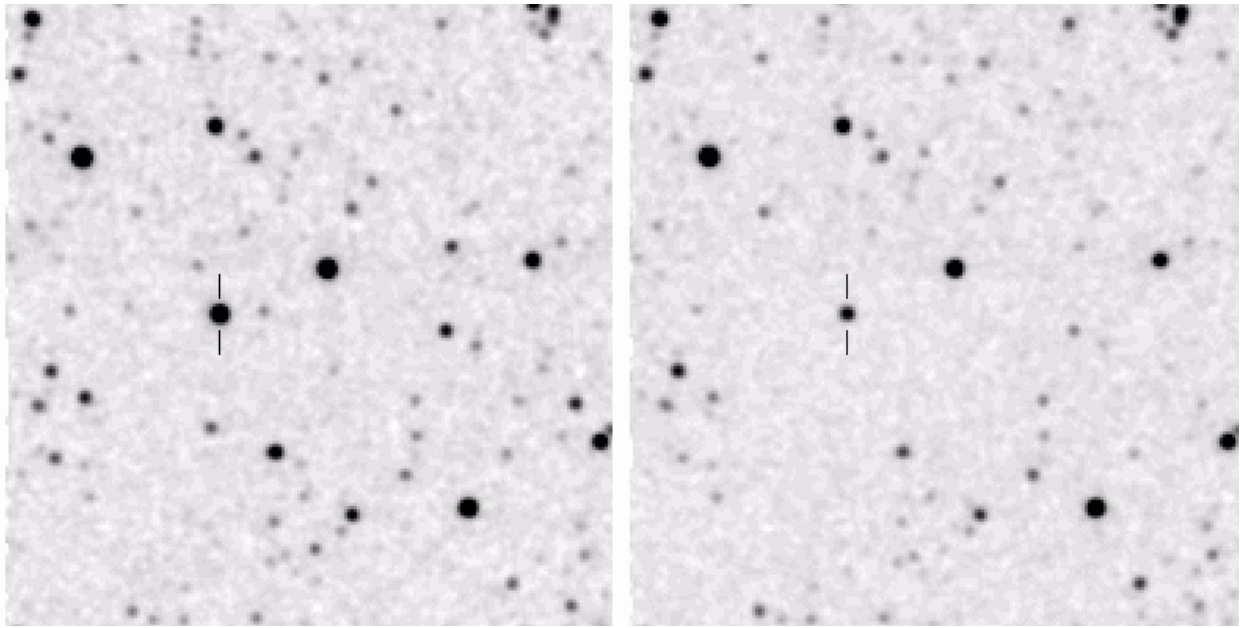

Figure 7. Infrared 2MASS images of DDE 24 obtained on 2000 May 24. Left: *K* band. Right: *J* band. Position of DDE 24 is marked with dashes. Note the high reddening around the variable.

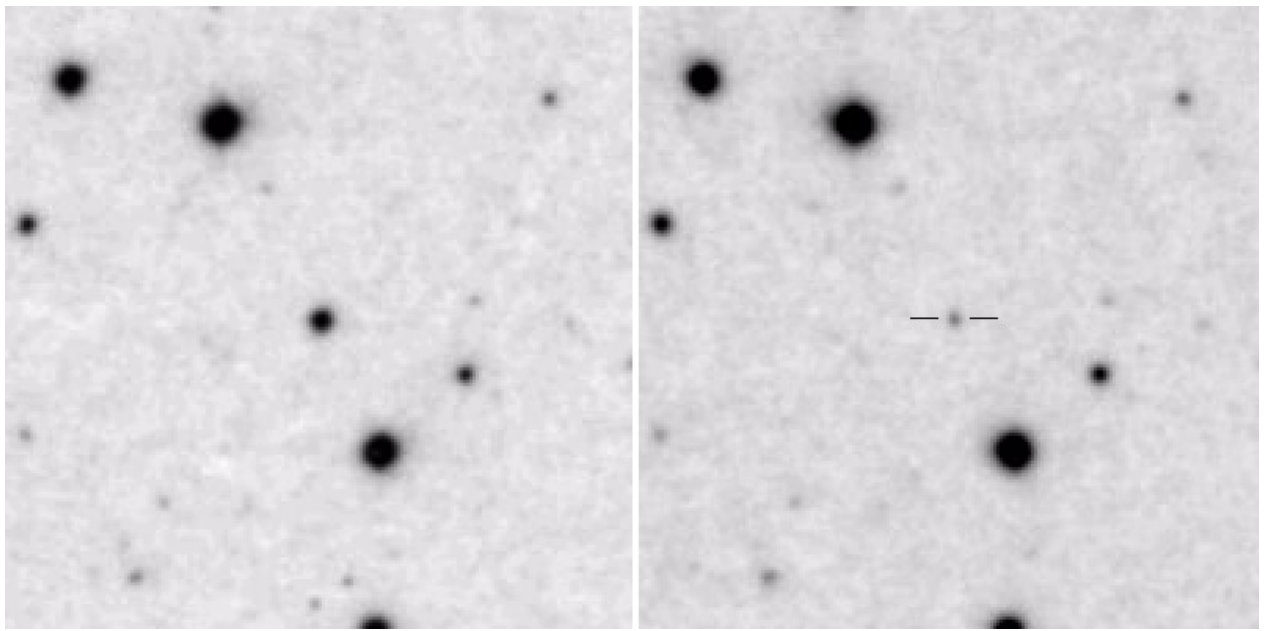

Figure 8. Blue Palomar images of DDE 25. Left: 1951 July 08 plate. Right: 1954 July 05 plate.

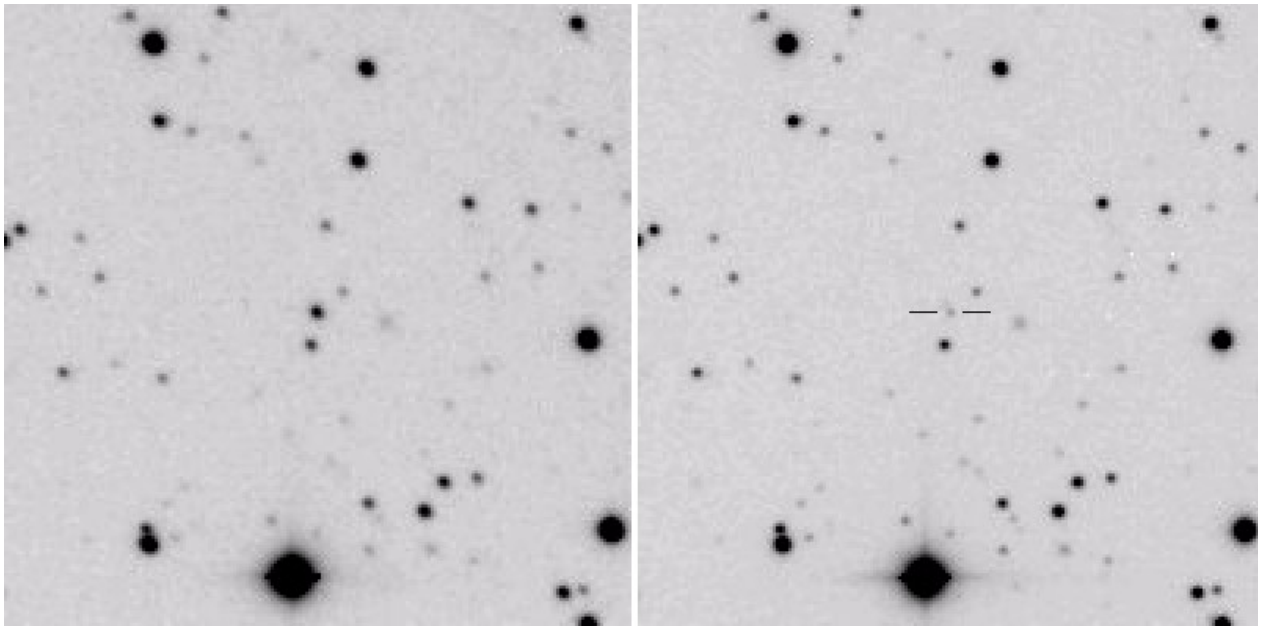

Figure 9. NEAT images of DDE 26 obtained in outburst (left) and at quiescence (right). Field of view is 200″×200″.

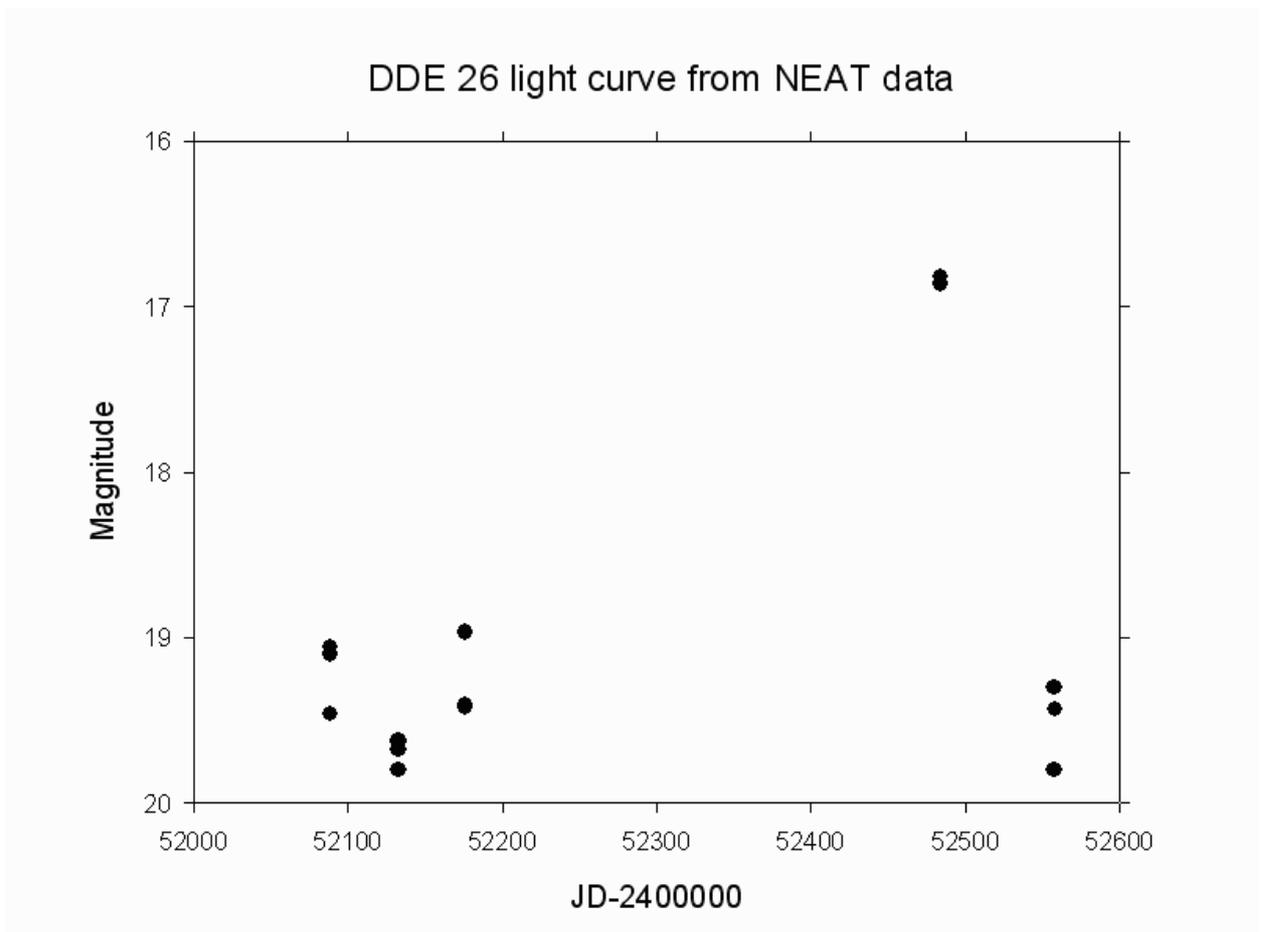

Figure 10. Light curve of DDE 26 from NEAT data.

13